\documentclass[aps,prb,twocolumn,superscriptaddress,showpacs]{revtex4-2}
\usepackage[dvipdfmx]{graphicx}
\usepackage{bm}
\usepackage{amsmath}
\usepackage{amssymb}
\usepackage{textcomp}
\usepackage{graphicx}
\usepackage{color}
\usepackage{subfigure}
\usepackage{soul,xcolor}
\usepackage{braket} 

\input{epsf}

\begin{document}
	
\title{Flux qubit-based detector of microwave photons}

\author{O.~A.~Ilinskaya}
\email{ilinskaya@ilt.kharkov.ua}
\affiliation{B.~Verkin Institute for Low Temperature Physics and Engineering, Kharkiv 61103, Ukraine}
\affiliation{G.~V.~Kurdyumov Institute for Metal Physics, Kyiv 03142, Ukraine}

\author{A.~I.~Ryzhov}
\affiliation{B.~Verkin Institute for Low Temperature Physics and Engineering, Kharkiv 61103, Ukraine}
\affiliation{Theoretical Quantum Physics Laboratory, Cluster for Pioneering Research, RIKEN, Wakoshi, Saitama, 351-0198, Japan}

\author{S.~N.~Shevchenko}
\affiliation{B.~Verkin Institute for Low Temperature Physics and Engineering, Kharkiv 61103, Ukraine}

\date{\today}

\begin{abstract}
A theory of detection of microwave photons with a flux qubit-based detector is presented. We consider semiclassical approximation with the electromagnetic field being in a coherent state. Flux qubit is considered as a multilevel quantum system (qu{\it d}it). By solving the Lindblad equation, we describe the time evolution of occupations of the qudit’s levels for readout and reset stages of detection. When considering the reset stage, the time evolution is described by multiple avoided-level crossings, thus providing a multilevel Landau-Zener-St\"uckelberg-Majorana (LZSM) problem. In addition to numerical calculations, we present an approximate solution for the description of the reset stage dynamics based on the adiabatic-impulse approximation and rate equation approach. Our theory may be useful for the theoretical description of driven-dissipative dynamics of qu{\it d}its, including applications such as single-photon detection.
\end{abstract}

\maketitle

\section{Introduction}

Detection of microwave photons is used in various fields of physics, for example, for measuring superconducting qubits \cite{opremcak2018, opremcak2021}, making two remote superconducting qubits entangled \cite{campagne-ibarcq}, imaging with a small number of photons \cite{morris}, and searching dark matter axions \cite{irastorza}. Different studies suggest different solutions for overcoming the difficulties in detecting photons of the microwave range which are because of small energies of these photons, and there is still no prevailing technology. Some of detectors of microwave photons are based on photon-assisted tunneling in semiconducting double quantum dot circuits (see Ref.~\cite{ghirri} for a review). Another direction of research uses superconducting circuits with Josephson junctions, which are nonlinear inductors and allow making superconducting qubits \cite{gu}. Usual photon detectors are destructive in the sense that a photon is absorbed making a transition of the qubit from the ground state to the excited one. This transition induces a ``click" of the detector. In addition, quantum non-demolition detection of microwave photons was reported \cite{kono, besse2018}.

Diverse types of superconducting qubits are used for detection of microwave photons --- phase qubits \cite{chen, opremcak2018}, flux qubits \cite{inomata}, transmons \cite{kono, besse2018}. 
A phase qubit is characterized by a ``washboard" potential and a ``click" means that a voltage is measured which is due to tunneling from the excited state of the qubit to the continuum. Tunneling from the ground state of the qubit through the barrier is also possible but the tunneling rate for this process is usually two-three orders of magnitude less than that of the excited state~\cite{chen}. However, this tunneling should be also taken into account and is responsible for dark count rate (when a ``click" happens without any photon). A phase qubit used in the detection scheme is called a Josephson photomultiplier \cite{govia} (see also Ref.~\cite{semenov}). Optimal characteristics for a phase qubit used in quantum computing and in photodetection are different. While quantum computers need long decoherence times to operate, for photodetection a dephasing time should be sufficiently small as was argued in Ref.~\cite{govia}.

In Ref.~\cite{chen}, a Josephson junction is galvanically coupled with control and readout circuits, and a ``click" of the detector is a voltage pulse. Because of the dissipative processes in the Josephson junction, Joule heat is released. One needs sufficiently long time for the system cooling, during which it is impossible to conduct the next measurement. The advantage of a flux qubit-based detector is that a Josephson junction is included into a superconducting loop and is isolated from control and readout circuits.

For a phase qubit, ``working" levels are localized in one well. On the contrary, for a flux qubit these two levels belong to different potential wells (i.e., they are characterized by opposite directions of a magnetic flux, or, which is the same, by opposite direction of supercurrent flowing in the loop). Therefore, a photon arrival changes the direction of the magnetic flux, and that is measured. A flux qubit-based photon detector was studied both theoretically \cite{koshino} and experimentally \cite{inomata}. In these papers, the dressed states of a superconducting qubit represented an artificial $\Lambda$-type three-level system. The signal pulse was in a weak coherent state with mean photon number of order 0.1. In the experiment, the signal-pulse shape was a Gaussian function (however, the authors theoretically confirm the possibility to detect signals with rectangular and exponentially-decaying shapes). The parameters which correspond to near-perfect absorption condition were found. At these  values of parameters, the reflection of the input signal is reduced. 

In this paper, we study a flux qubit-based detector of microwave photons in a semiclassical approximation, which means that the input signal is in a coherent state. No additional driving fields, used in Ref.~\cite{koshino} to create the dressed qubit-resonator system, are needed in our model. Unlike Ref.~\cite{koshino}, we consider the flux qubit as a multilevel system. 
The potential energy and three levels, on which the detection mechanism is based (see  Fig.~\ref{Fig:PrincipleOfOperation} below), should satisfy the following conditions: (i)	energy difference between two ``working" levels corresponds to the frequency of external microwave field; (ii) the barrier height allows tunneling from the upper ``working" level to lower levels and blocks thermally activated transitions above the barrier; (iii) the tunneling probability from the lower ``working" level is extremely low; (iv) the deep well is sufficiently deep so that the change of magnetic flux, which occurs during the readout stage, is fixed for a long time. We include into consideration several additional levels, which appear in the deep well due to these conditions.
We note that some experimental studies of a flux qubit-based photon detector of this kind were presented in Refs.~\cite{shapovalov, lyakhno} and the scheme of weak continuous measurement for readout of the flux qubit states was proposed~\cite{shnyrkov2020}. Using numerical calculations, we plot the dynamics of qubit's populations for readout and reset stages of detection.

The paper is organized as follows. In Section~\ref{Sec:Superconducting flux qubit: electric circuit, Lagrangian, stationary Hamiltonian} we write down the stationary Hamiltonian of a flux qubit and obtain stationary energy levels and eigenfunctions numerically. In Section~\ref{Sec:Principle of operation of the photon detector} we discuss the principle of operation of the detector. Section~\ref{Sec:Dynamics} is devoted to the calculation of the dynamics of the occupations of the levels of the multilevel system for readout and reset stages. We conclude in Section~\ref{Sec:Conclusions}. Appendix~A contains the approximate description of the reset stage dynamics.

\section{Superconducting flux qubit: electric circuit, Lagrangian, stationary Hamiltonian}
\label{Sec:Superconducting flux qubit: electric circuit, Lagrangian, stationary Hamiltonian}

\begin{figure}
	\includegraphics[width=5cm]{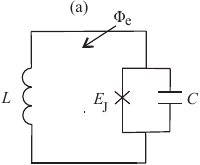}
	
	\vspace{0.5cm}
	\includegraphics[width=8cm]{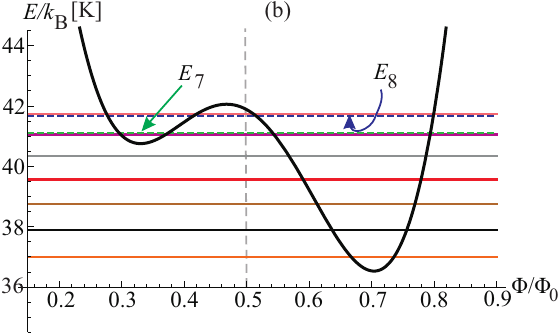}
	
	\vspace{0.5cm}
	\includegraphics[width=8cm]{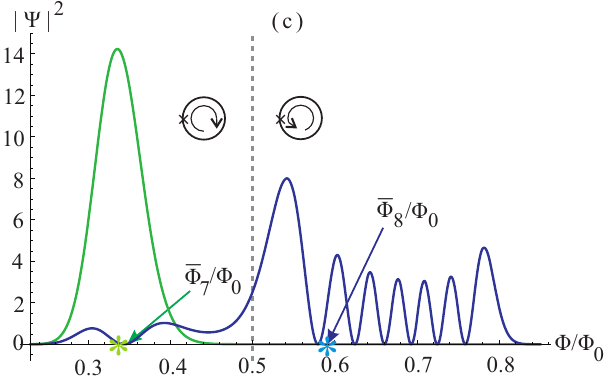}
	\caption{(a) The electric circuit describing a superconducting flux qubit. $E_\text{J}$ is the Josephson energy, $L$ is the inductance, $C$ is the capacitance, $\Phi_\text{e}$ is the external magnetic flux. (b) The dependence of the potential energy~(\ref{potential}) and the qudit levels' energies (in kelvins) on the magnetic flux $\Phi$ (normalized to the magnetic flux quantum $\Phi_0$). There are nine levels localized in one of the two wells. The levels, which are in resonance with incoming microwave pulse, are shown by the dashed lines (the 7th and 8th levels). Note that in the selected scale, the 6th and 7th levels are practically indistinguishable to the eye (and the same is true for the 8th and 9th levels). The proximity of the 6th and 7th (and the 8th and 9th) levels is due to splitting of levels caused by the tunneling through the barrier.  (c)~The square of the eigenfunction absolute value as a function of $\Phi/\Phi_0$ for the 7th level (the green curve) and the 8th level (the blue curve). The asterisks on the $x$-axis denote the average values of the full magnetic flux for the two levels. Arrows inside qubit's loop show the direction of the electric current. The simulation parameters are taken close to the ones of Ref.~\cite{shnyrkov}: magnetic energy $U_0=32.68\,\text{K}$,  dimensionless inductance $\beta_L=1.28$,  effective mass $M=955\,\text{K}^{-1}$, external magnetic flux (in units of $\Phi_0$) $x_\text{e}=0.5087$.}
	\label{Fig:EnergiesAndWaveFunctions}
\end{figure}

A superconducting flux qubit can be described by the following electric circuit [see Fig.~\ref{Fig:EnergiesAndWaveFunctions}(a)]: a loop, which contains a Josephson junction and an inductance $L$ and is pierced by an external magnetic flux $\Phi_\text{e}$. The full magnetic flux $\Phi$ in the loop and $\Phi_\text{e}$ are related by the transcendental equation 
\begin{equation}\label{MagneticFluxes}
\Phi=\Phi_\text{e} - L I_\text{c}\sin\left(2\pi\frac{\Phi}{\Phi_0}\right),
\end{equation}
which is due to the existence of a shielding current and holds for radio-frequency SQUID \cite{schmidt}; here 
$I_\text{c}$ stands for the critical current, $\Phi_0=\pi\hbar/e$ is the magnetic flux quantum. 
The energy associated with the capacitance $C$, $C\dot{\Phi}^2/2$, can be considered as a kinetic energy. There are two potential energy terms, i.e. $(\Phi-\Phi_\text{e})^2/2 L$ associated with the inductance and the Josephson junction energy $E_\text{J}[1-\cos(2\pi\Phi/\Phi_0)]$, with $E_\text{J} = \Phi_0 I_\text{c} / 2\pi$. Then the Lagrangian can be written as 
\begin{equation}
{\cal L}=\frac{C\dot{\Phi}^2}{2} - \frac{(\Phi-\Phi_\text{e})^2}{2 L} - 
E_\text{J}\left[1 - \cos\left(2\pi\frac{\Phi}{\Phi_0}\right)\right].
\end{equation}
The variable canonically conjugated to the flux $\Phi$ is defined as $Q = \partial{\cal L}/\partial\dot{\Phi} = C \dot{\Phi}$. Treating magnetic flux $\Phi$ and charge $Q$ quantum-mechanically, the charge operator can be expressed as $Q = -i\hbar\,\partial/\partial\Phi$. 

Omitting the constant term, the (stationary) Hamiltonian can be written as \cite{shnyrkov} 
\begin{equation}\label{stationary}
H(\Phi, Q)=\frac{Q^2}{2 C}+U(\Phi,\Phi_\text{e}),
\end{equation}
where the potential energy is given by
\begin{equation}
U(\Phi,\Phi_\text{e})=-E_\text{J}\cos\left(2\pi\frac{\Phi}{\Phi_0}\right)+
\frac{(\Phi-\Phi_\text{e})^2}{2 L}.
\end{equation}
The stationary Schr\"odinger equation with energies measured in kelvins has the form 
\begin{equation}\label{Schroedinger}
\left\{-\frac{1}{2 M}\frac{\partial^2}{\partial x^2} + U(x)\right\}\Psi(x)=
E\Psi(x),
\end{equation}
with the potential energy
\begin{equation}\label{potential}
U(x)=U_0 \left\{-\beta_L \cos(2 \pi x) + 2 \pi^2 (x - x_\text{e})^2\right\}.
\end{equation}
Here 
\begin{equation}
x=\Phi/\Phi_0, \qquad x_\text{e}=\Phi_\text{e}/\Phi_0,
\end{equation}
and
\begin{equation}
U_0=(\Phi_0/2\pi)^2/k_\text{B} L, \qquad M = k_\text{B} \Phi_0^2 C/\hbar^2,
\end{equation}
with $U_0$ and  $M$ being the magnetic energy and the effective mass  ($k_B$ is the Boltzmann constant); 
\begin{equation}
\beta_L = 2 \pi L I_\text{c} / \Phi_0,
\end{equation} 
is the dimensionless inductance. 

The parameter $\beta_L$ is a tunable quantity due to controllable critical current. (In practice, this tunability is realized by replacing one Josephson junction with a small loop containing two junctions.) From the analysis of Eq.~(\ref{potential}), it follows that $U(x)$ is a two-well potential if the double inequality holds $1/\pi < \beta_L < 2.48$, and we take $\beta_L$ from this interval below. Parameters $U_0$ and $M$ are determined by the inductance $L$ and capacitance $C$ respectively, therefore, they are determined by the qubit design. 

Equation~(\ref{Schroedinger}) has a formal analogy with the Schr\"odinger equation for a quasiparticle with the effective mass $M$ moving in the potential $U(x)$. The eigenenergies of the Hamiltonian~(\ref{stationary}) can be found numerically \cite{shnyrkov} and, for the parameters of Fig.~\ref{Fig:EnergiesAndWaveFunctions} (close to those of Ref.~\cite{shnyrkov}), there are nine energy levels localized in one of the wells. These levels are shown in Fig.~\ref{Fig:EnergiesAndWaveFunctions}(b). [We say that the level is localized in a well if its eigenfunction $\Psi_n(x)$ is localized in this well, see Fig.~\ref{Fig:EnergiesAndWaveFunctions}(c)]. Levels localized in different  wells correspond to opposite directions of the electric current. Indeed, the external magnetic flux generates shielding current $I_\text{s}=-I_\text{c}\sin(2\pi\Phi/\Phi_0)$ in the loop. Therefore, $I_\text{s}<0$ for $\Phi/\Phi_0<0.5$ and $I_\text{s}>0$ for $\Phi/\Phi_0>0.5$.

\section{Principle of operation of the photon detector}
\label{Sec:Principle of operation of the photon detector}

\begin{figure}[t] 
	\includegraphics[width=8.5cm]{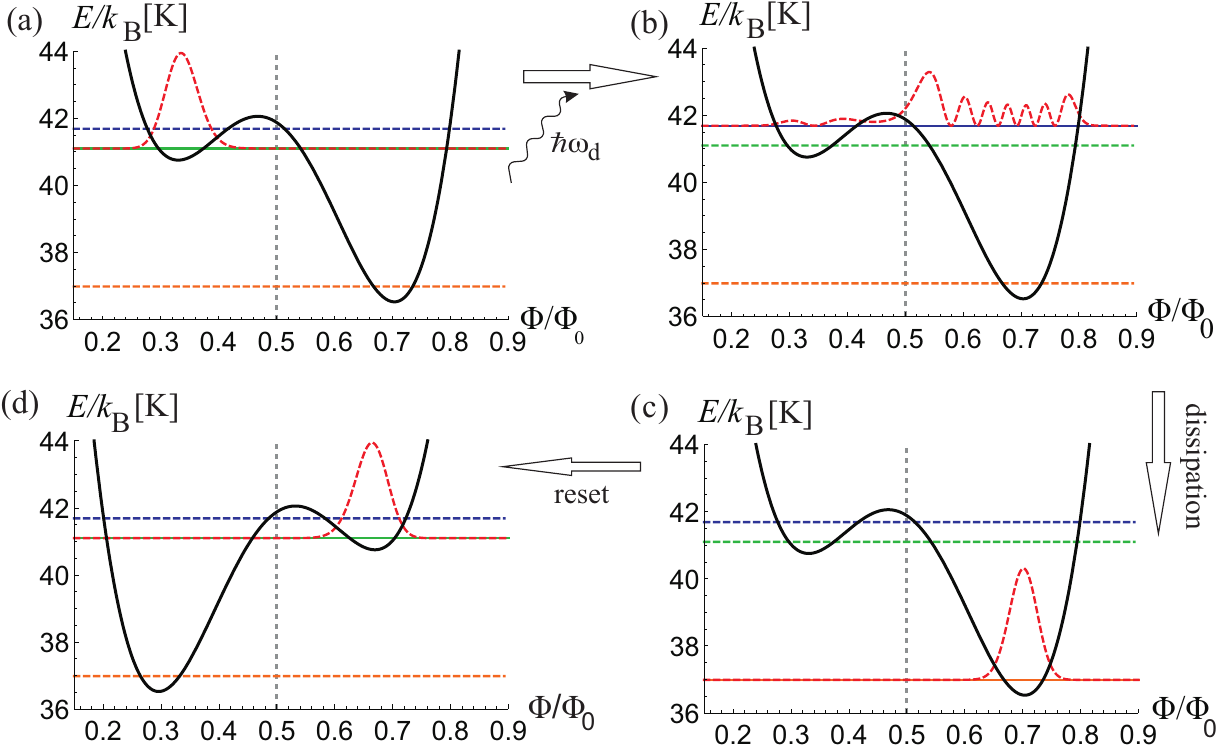}
	\caption{ The principle of operation of the photon detector. (a) The quasiparticle is on the 7th level [only the levels Nos.~1, 7, and 8 are shown, cf.~Fig.~\ref{Fig:EnergiesAndWaveFunctions}(b)]. (b) The incoming signal with the frequency $\omega_\text{d}=(E_8 - E_7)/\hbar$ induces the Rabi oscillations between the 7th and 8th levels (the capture stage of the detection). (c) The dissipation transfers the quasiparticle to the lowest level, therefore, the change of the direction of the electric current can be measured (the readout stage of the detection). (d) Reset of the system is done by changing the external magnetic flux $\Phi_\text{e}$ from the value 0.5087 to 0.4913 [see also Fig.~\ref{Fig:E_Phi-e}(a)]. The potential $U(x)$ is distorted and the quasiparticle is on the 7th level again. }
	\label{Fig:PrincipleOfOperation}
\end{figure}

Principle of operation of the flux qubit-based photon detector is shown in Fig.~\ref{Fig:PrincipleOfOperation}.
Let the quasiparticle be on the 7th level. The incoming microwave pulse (which should be detected) is resonant with the pair of levels, namely levels 7 and 8, which are localized in the left and right wells respectively, as it can be seen from Fig.~\ref{Fig:EnergiesAndWaveFunctions}(b,~c). These levels' energies are shown by the dashed lines in Fig.~\ref{Fig:EnergiesAndWaveFunctions}(b). The electromagnetic pulse induces the Rabi oscillations between the 7th and 8th levels. Due to dissipation, the quasiparticle falls down to the lowest level in the right well, therefore,  the electric current in the loop changes its direction that corresponds to signal detection. It is necessary to distort the form of the potential (by changing the external magnetic flux $\Phi_\text{e}$) in order to make the right well shallow and quasiparticle being on the 7th level so that the system is ready to register another signal. 

Note that the Rabi oscillations should be between the levels, which are localized in different wells, as the magnetic flux should substantially change during the detection process (the readout stage). Also, during the detection process, the quasiparticle should go from the shallow well into the deep one in order to have the change of magnetic flux fixed. These two conditions in our case (with the parameters taken) imply that the ``working" levels are the 7th and 8th ones.

\begin{figure} 
	\includegraphics[width=8.0cm]{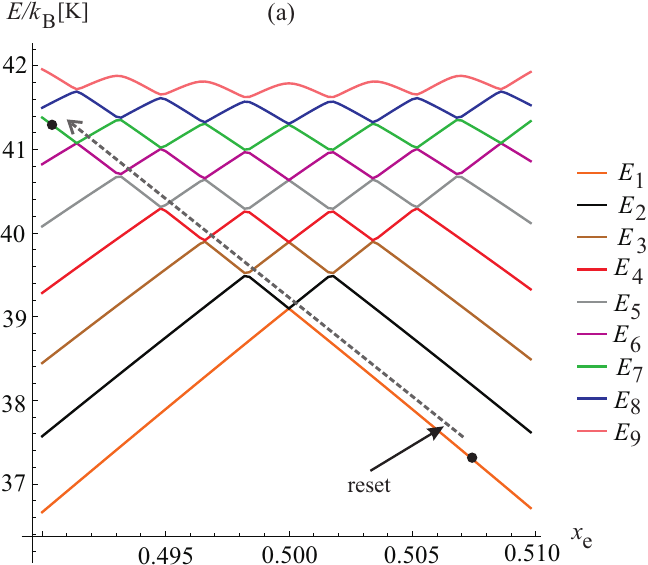}
	
	\vspace{0.5cm}
	\hspace{-0.8cm}\includegraphics[width=7.8cm]{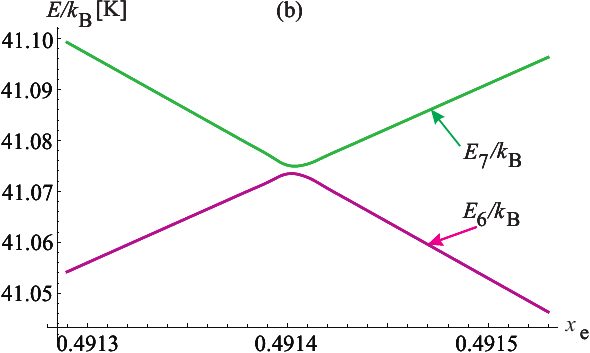}
	
	\vspace{0.5cm}
	\includegraphics[width=8.0cm]{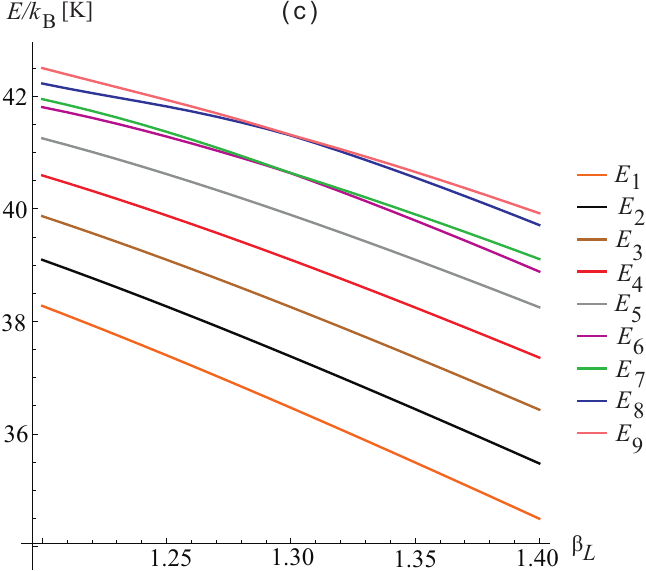}
	\caption{(a) The qudit energy levels as a function of the external magnetic flux $x_\text{e}$. The energy levels exhibit quasicrossings as it is shown in panel~(b) for the 6th and 7th levels. The simulation parameters are the same as in Fig.~\ref{Fig:EnergiesAndWaveFunctions}. (c) The dependence of the qudit energy levels on the dimensionless inductance $\beta_L$. The simulation parameters are $U_0 = (41.67 / \beta_L)\,\text{K}$, $M=955\,\text{K}^{-1}$, $x_\text{e}=0.5087$. }
	\label{Fig:E_Phi-e}
\end{figure}

The dependence of the levels' energies on the external magnetic flux $\Phi_\text{e}$ is shown in Fig.~\ref{Fig:E_Phi-e}(a). The avoided crossing, inherent to all pairs of adjacent levels, is illustrated for the 6th and 7th levels in Fig.~\ref{Fig:E_Phi-e}(b). Fig.~\ref{Fig:E_Phi-e}(c) shows the levels' energies as functions of one of the system parameters, $\beta_L$.

We now estimate the speed of external magnetic flux variation which ensures a transition from the lowest level to the 7th level during the reset with sufficiently high transition probability $(P\approx 0.99)$. This speed $v$ (in energy units) can be estimated from the Landau-Zener-St\"uckelberg-Majorana formula (see, for example, Refs.~\cite{ashhab, kofman, ivakhnenko}). This formula gives the probability of diabatic transition between two energy states of a quantum system which has a time-dependent Hamiltonian with linear bias, i.e.
\begin{equation}
H_{\text{LZSM}} = \frac{1}{2} \left( \begin{array}{cc} v t & \Delta \\ \Delta & - v t \end{array} \right).
\end{equation}
If the system starts, in the infinite past, in the lower energy eigenstate, the probability of finding the system in the upper energy eigenstate in the infinite future, after traversing the avoided crossing region, is as follows
\begin{equation}\label{final_occupation}
P_{\text{LZSM}}=e^{-\frac{\pi\Delta^2}{2\hbar v}}.
\end{equation}
Now, we have a task with fixed target $P_{\text{LZSM}}$ and several values of $\Delta$'s, see Fig.~\ref{Fig:E_Phi-e}(a). Then, the minimal speed $v$, which is needed to achieve the target transition probability, corresponds to the largest avoided-level crossing $\Delta_\text{max}\approx 3\,\text{mK}$. The relation between $v$ and the speed $d\Phi_\text{e}/dt$ of external magnetic flux variation is the following \cite{omelyanchouk}: $d\Phi_\text{e}/dt=v/2I_\text{p}$, where we take the persistent current $I_\text{p}\approx\Phi/L\approx 3\mu\text{A}$. Using this estimate and the value of $v$ corresponding to $P_{\text{LZSM}}=0.99$, we obtain $d\Phi_\text{e}/dt\approx 0.5~\text{nWb/s}\approx 0.25 \Phi_0/{\mu\text{s}}$ (note that this estimate is obtained in the absence of dissipation). This is a realistic value for rf SQUIDs. Varying $\Phi_\text{e}$ with this speed, we need approximately $0.1\,\mu\text{s}$ to change $\Phi_\text{e}$ from the initial value 0.5087$\Phi_0$ to the final value 0.4913$\Phi_0$. We compare our estimate of the reset time with the values mentioned in the papers studying the detectors based on a phase qubit. We note the value of 1~ms given in Ref.~\cite{shnyrkov} and the one of 0.1~$\mu\text{s}$ in Ref.~\cite{govia2014}. 
Our estimate is much less than the corresponding value in Ref.~\cite{shnyrkov} and of the same order as the one in Ref.~\cite{govia2014}.

\section{Dynamics}
\label{Sec:Dynamics}

\subsection{Capture and readout stages}

\subsubsection{Full Hamiltonian and semiclassical approximation}

We write the Hamiltonian~(\ref{stationary}) in the basis of eigenfunctions of the qudit. Taking also into account the single-mode resonator and incoming microwave pulse,
which is in resonance with the 7th and 8th levels of the qudit, 
we can write down the Hamiltonian as follows \cite{brookes}
\begin{eqnarray}\label{Jaynes-Cummings}
H &=& \sum_{j=1}^{N}E_j |E_j\rangle \langle E_j|+
\hbar\omega_\text{r} a^\dag a +
g (a^\dag |E_7\rangle \langle E_8| + \text{h.c.}) \nonumber\\ 
&+& 
A_\text{d} f(t) (a^\dag e^{-i\omega_\text{d} t} + \text{h.c.}).
\end{eqnarray}
This is the simplification of the generalized Jaynes-Cummings Hamiltonian, $N=9$ for the chosen parameters. By simplification, we mean that the interaction of the qudit with the electromagnetic field, described by the third term of Eq.~(\ref{Jaynes-Cummings}), couples the field just to the 7th and 8th levels and leaves all other levels unaffected. Here $a$ is the annihilation operator for electromagnetic field, $E_j$ is the eigenenergy of the $j$th level corresponding to the eigenfunction $|E_j\rangle$ of the qudit, $\omega_\text{r}$ is the resonator frequency, $g$ is the qudit-field interaction energy, $A_\text{d}$ and $\omega_\text{d}$ are the energy and the angular frequency of the incoming microwave pulse respectively. The signal can have almost any shape $f(t)$, for example, Gaussian, square, exponential \cite{koshino}. 
Note that the Hamiltonian of a flux qubit with a single Josephson junction inductively coupled to a resonator has the form of the quantum Rabi Hamiltonian \cite{rabi, jaynes-cummings} if we do not include into consideration the qubit's higher energy levels \cite{yoshihara2022}. We note also that energy eigenstates of a flux qubit--resonator system with deep strong coupling were reported to be entangled ones \cite{yoshihara2017}.

We perform a unitary rotating-wave transformation, 
\begin{equation}
U = \exp[i\omega_\text{d} t (a^\dag a+\sum_j j |E_j\rangle \langle E_j|\,)]
\end{equation} 
(see, for example, Ref.~\cite{baker}), after which the transformed Hamiltonian, $\tilde H = U H U^\dag + i\hbar\dot U U^\dag$, takes the form 
\begin{eqnarray}\label{tildeH}
	\tilde H &=& \sum_{j=1}^{N}(E_j - \hbar\omega_\text{d}j) |E_j\rangle \langle E_j|+
	\hbar(\omega_\text{r}-\omega_\text{d}) a^\dag a \nonumber\\
	&+&
	g (a^\dag |E_7\rangle \langle E_8| + \text{h.c.})  
	+ 
	A_\text{d} f(t) (a^\dag + a).
\end{eqnarray}
We assume that the field is in the coherent state $| \alpha \rangle$, average $\tilde H$ over $| \alpha \rangle$ analogously to Ref.~\cite{shevchenko}, and obtain $H_{\text{c}} = \langle \alpha | \tilde H | \alpha \rangle$ in the form
\begin{eqnarray}\label{Hc}
	H_{\text{c}} &=& \sum_{j=1}^{N}(E_j - \hbar\omega_\text{d}j) |E_j\rangle \langle E_j|+
	\hbar(\omega_\text{r}-\omega_\text{d}) \langle n \rangle \nonumber\\
	&+&
	g (\alpha^\ast |E_7\rangle \langle E_8| + \text{h.c.})  
	+ 
	A_\text{d} f(t) (\alpha^\ast + \alpha),
\end{eqnarray}
where $\alpha = \sqrt{\langle n \rangle}$ and $\langle n \rangle$ is the mean photon number. Assuming for simplicity $\alpha$ being real and omitting constants, we can rewrite Eq.~(\ref{Hc}) as follows
\begin{eqnarray}\label{Hc2}
	H_{\text{c}} &=& \sum_{j=1}^{N}(E_j - \hbar\omega_\text{d}j) |E_j\rangle \langle E_j|\nonumber\\
	&+&
	g \alpha (\,|E_7\rangle \langle E_8| + \text{h.c.}) 
	+ 
	2 A_\text{d} \alpha f(t).
\end{eqnarray}

\begin{figure}[t]
	\includegraphics[width=8.4cm]{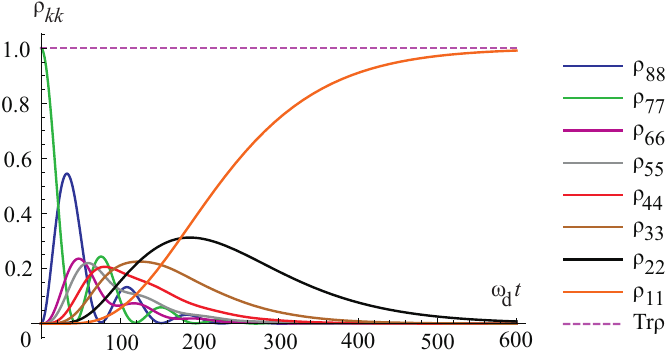}
	\caption{ Time dependence of the occupations of the qudit levels for the readout stage.  The simulation parameters are $U_0=32.68\,\text{K}$, $\beta_L=1.28$, $M=955\,\text{K}^{-1}$, $x_\text{e}=0.5087$,  $g=2.5\cdot 10^{-3}\,\text{K}$, $\alpha=10$, the dissipation rate is $\gamma=2\cdot 10^{-5}\,\text{K}$ (approximately 0.4\,\text{MHz}). The frequency of external microwave pulse is $\omega_\text{d}=(E_8-E_7)/\hbar$. }
	\label{Fig:rho_t}
\end{figure}

\subsubsection{Lindblad equation and qudit's populations dynamics}

For the capture stage (excitation) and readout stage (relaxation), the Lindblad equation for the density operator $\hat{\rho}$ reads  
\begin{equation}\label{LindbladEquation}
\partial_t\hat{\rho} = -\frac{i}{\hbar} [H_{\text{c}},\,\hat{\rho}]+\sum_{n,n' (n>n')}\Gamma_{nn'}D(|n'\rangle\langle n|)\hat{\rho}.
\end{equation}
Here we take into account the relaxation in the qudit analogously to Ref.~\cite{jones} and neglect dissipation in the resonator which is assumed to be small \cite{govia}. Superoperator $D(O)\hat{\rho}$ has the standard form \cite{shevchenko} 
\begin{equation}
	D(O)\hat{\rho}=O\hat{\rho} O^\dag - \frac{1}{2}O^\dag O\hat{\rho}
	- \frac{1}{2}\hat{\rho} O^\dag O,
\end{equation}
and the relaxation rates are given by 
\begin{equation}\label{relaxation-rates}
\Gamma_{nn'}=\gamma\frac{E_n-E_{n'}}{\hbar}\Big|\Big\langle \Psi_{n'}(x)\Big|\frac{\partial}{\partial x}\Psi_n(x)\Big\rangle\Big|^2 \theta(n-n').
\end{equation}
Here $\gamma$ is a dimensionless overall factor, the derivative is due to the charge operator $\hat{Q}$, and $\theta$ stands for the Heaviside step function. The temperature dependence of the relaxation rates can be included (see, for example, Ref.~\cite{makhlin}) by multiplying the last term in the r.h.s. of Eq.~(\ref{LindbladEquation}) by a function $\coth[(E_n-E_{n'})/2 k_\text{B} T]$. However, this factor can be neglected for temperatures $T\sim 20\,\text{mK}$~\cite{shnyrkov}. Note that we do not consider the pure dephasing here, although there are many papers investigating different sources and types of the noise leading to the dephasing in superconducting qubits. We emphasize that white noise (with the power spectral density linear in temperature) is small at low temperatures, while low-frequency flux noise should be taken into account~\cite{kumar}. Several experimental works show that $1/f$ flux noise is a dominant source of the dephasing in flux qubits (see, for example, Ref.~\cite{yoshihara2006}). In Ref.~\cite{rower}, the white noise floor is present in addition to $1/f$ noise.

The differential equation for the matrix element $\rho_{kk'}\equiv\langle E_k | \hat{\rho} | E_{k'} \rangle$, taking Eq.~(\ref{LindbladEquation}) into account, has the form 
\begin{align}\label{system2}
	&\frac{d\rho_{kk'}}{dt}=
	-i\Big\{\left(\frac{E_k}{\hbar}-\omega_\text{d} k\right)\rho_{kk'}-
	\left(\frac{E_{k'}}{\hbar}-\omega_\text{d} k'\right)\rho_{kk'}\nonumber\\
	&+\frac{g}{\hbar}\alpha
	(\rho_{8k'}\delta_{k7}+\rho_{7k'}\delta_{k8}-
	\rho_{k7}\delta_{k'8}-\rho_{k8}\delta_{k'7})\Big\}\nonumber\\
	&+\sum_{j=1}^8\left[\Gamma_{jk}\rho_{jj}\delta_{kk'}
	-\frac{1}{2}(\Gamma_{kj}+\Gamma_{k'j})\rho_{kk'}\right],
\end{align}
where $\delta_{kk'}$ is the Kronecker symbol. 
Note that the form of the signal $f(t)$ does not enter Eq.~(\ref{system2}) in the semiclassical approximation. The system of equations (\ref{system2}) for $k, k'=1, 2, ..., 8$ is solved numerically.

Temporal evolution of the diagonal matrix elements $\rho_{kk}$ 
for the readout stage 
is shown in Fig.~\ref{Fig:rho_t}. For $t\to\infty$, the occupation of the lowest level tends to unity while occupations of all other levels vanish. There are damped Rabi oscillations between levels 7 and 8. An increase of $g$ (or $\alpha$) leads to the growth of the frequency of these oscillations and vice versa. (This statement is not illustrated here.) 

\subsection{Reset stage}

In order to obtain time evolution of the occupations of the qudit's energy levels during the reset stage, it is necessary to solve the Lindblad equation~(\ref{LindbladEquation}) with $H_\text{c}$ from Eq.~(\ref{Hc2}) replaced by the time-dependent Hamiltonian 
\begin{equation}
H_\text{r}(t)=\sum_{j=1}^{N}E_j(t) |E_j(t)\rangle \langle E_j(t)|.
\end{equation} 
Here the time dependence is due to variation of the external magnetic flux, 
\begin{equation}\label{Phi-e_reset}
\Phi_\text{e}=\Phi_{\text{e}0} - v_\text{e} t,
\end{equation} 
with $v_\text{e}=d\Phi_\text{e}/dt$, and $|E_j(t)\rangle$ is the eigenfunction corresponding to the $j$th adiabatic energy level $E_j(t)$. We denote via $\hat{A}(t)$ the transfer matrix from the eigenenergy basis, with $t=0$, to the adiabatic basis, then $A_{kk'}(t)=\langle E_{k'}^{(0)}|E_k(t)\rangle$, where $|E_{k'}^{(0)}\rangle=|E_{k'}(t=0)\rangle$. Note that the problem of reset is, in fact, the ``equal-slope" multilevel Landau-Zener-St\"uckelberg-Majorana problem (see Ref.~\cite{ivakhnenko} for a review).

\begin{figure}
	\includegraphics[width=7.5cm]{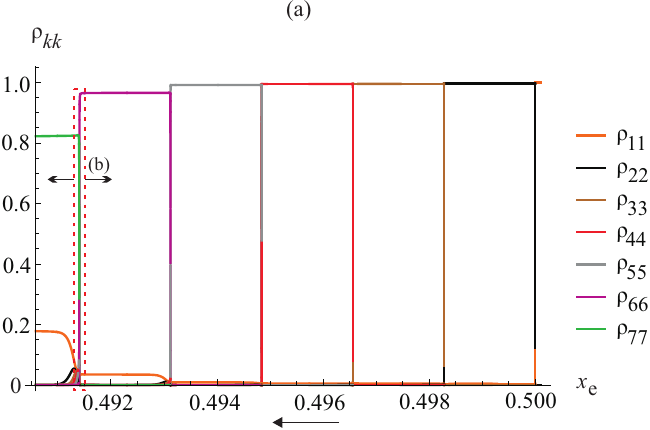}
	
	\vspace{0.3cm}
	\includegraphics[width=7.5cm]{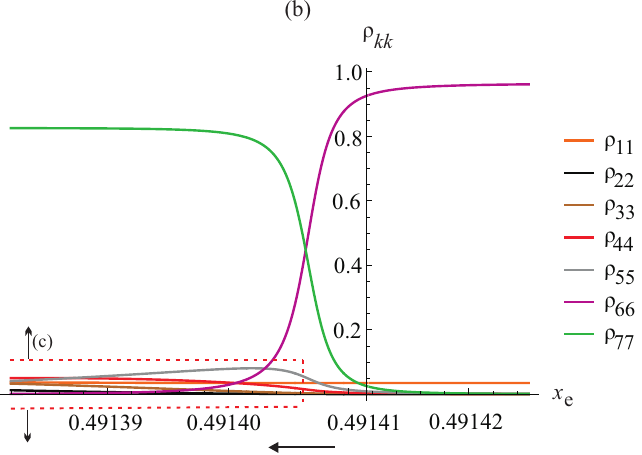}
	
	\vspace{0.3cm}
	\includegraphics[width=7.5cm]{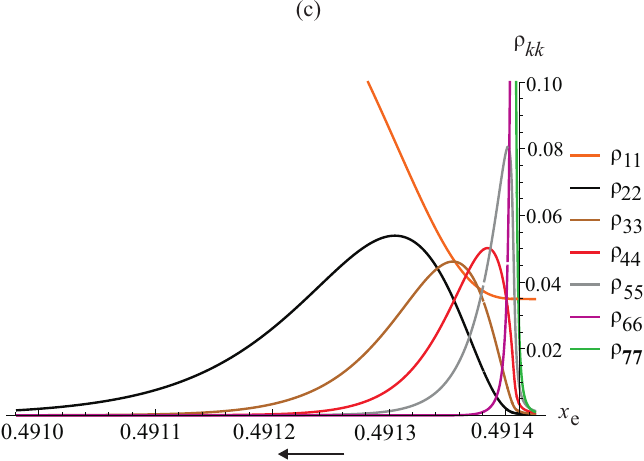}
	\caption{(a) The dependence of the occupations of the qudit levels on the dimensionless external magnetic flux $x_\text{e}$ that describes the reset stage of the detection. The change of $x_\text{e}$ is towards lower values (which is shown by arrows below the graphs). 
	The simulation parameters are $U_0=32.68\,\text{K}$, $\beta_L=1.28$, $M=955\,\text{K}^{-1}$, $\tilde{v}_\text{e}(=v_\text{e}/\Phi_0)=0.1\,\text{GHz}$, $\gamma=3\,\text{GHz}$. We take the initial value of external magnetic flux in Eq.~(\ref{Phi-e_reset}) as follows: $\Phi_{\text{e}0} = 0.5001\Phi_0$ instead of $0.5087\Phi_0$ (cf. Fig.~\ref{Fig:EnergiesAndWaveFunctions}) because the dynamics for the values of external magnetic flux larger than $0.5\Phi_0$ is trivial (there are no transitions). (b)~The same dependence as in~(a) near the point of quasicrossing of levels 6 (magenta) and 7 (green). The characteristic width of this transition in units of magnetic flux is about $10^{-5}\Phi_0$. (c)~The same dependence as in~(a) that shows the dynamics of occupations of intermediate levels (Nos.~5, 4, 3, 2) due to dissipation after the last quasicrossing point (zoom-in near the $x$-axis). Note that the growth of $\rho_{11}$ (the orange curve), present in~(c), is not seen in~(b) because of different ranges on the axes. }
	\label{Fig:Reset}
\end{figure}

Taking into account that 
\begin{align}\label{rho-dot}
\langle E_k(t) | &\frac{d\hat{\rho}}{dt} | E_{k'}(t) \rangle = 
\frac{d}{dt}\langle E_k(t) | \hat{\rho} | E_{k'}(t) \rangle \nonumber\\
&- \frac{d\langle E_k(t)|}{dt}\hat{\rho} | E_{k'}(t) \rangle - \langle E_k(t) | \hat{\rho} \frac{d | E_{k'}(t) \rangle}{dt}
\end{align}
and neglecting dephasing, 
we obtain the differential equation for the matrix element $\rho_{kk'}\equiv\langle E_k(t) | \hat{\rho} | E_{k'}(t) \rangle$ for the reset stage in the form
 \begin{align}\label{system-reset}
 &\frac{d\rho_{kk'}}{dt}=
 \Big(\hat{B}^\ast(t)\hat{\rho}\Big)_{kk'} + 
 \Big(\hat{B}(t)\hat{\rho}^\ast\Big)_{k'k}\nonumber\\
 &-\frac{i}{\hbar}\Big(E_k(t) - E_{k'}(t)\Big)\rho_{kk'}\nonumber\\
 &+\sum_{j=1}^7\left[\Gamma_{jk}(t)\rho_{jj}\delta_{kk'}
 -\frac{1}{2}(\Gamma_{kj}(t)+\Gamma_{k'j}(t))\rho_{kk'}\right],
 \end{align}
where $k, k'=1, 2, ..., 7$ and
\begin{equation}\label{B}
\hat{B}(t)=\frac{d\hat{A}(t)}{dt}\hat{A}^{-1}(t).
\end{equation} 
Note that the relaxation rates $\Gamma_{kk'}$ are now time-dependent since they include eigenenergies and eigenfunctions [see Eq.~(\ref{relaxation-rates})].

We plot the dependence of the occupations of the qudit levels on the external magnetic flux $x_\text{e}$ for the reset stage in Fig.~\ref{Fig:Reset}(a). Note that we take the speed $v_\text{e}$ of the external magnetic flux variation much larger than the minimal necessary value which we obtained as an estimate in Section~III. There is a technical reason for it as the larger the speed, the less the time of numerical calculations. Due to large value of $v_\text{e}$, the characteristic widths of the transitions are small and vary from $10^{-9}$ (in units of magnetic flux quantum) for the first transition to $10^{-5}$ for the last transition. 
Fig.~\ref{Fig:Reset}(b) shows a part of Fig.~\ref{Fig:Reset}(a) corresponding to LZSM transition from level 6 to level 7. This subfigure is obtained from Fig.~\ref{Fig:Reset}(a) by stretching the area bounded by the red dashed rectangle in Fig.~\ref{Fig:Reset}(a) along the abscissa axis. The probability $P_\text{r}$ to return to level 7 from level 1 is about 0.83 for chosen $\gamma$ [overall factor in Eq.~(\ref{relaxation-rates})]. We note that dissipation is almost negligible till the transition from level 5 to level 6 and that the dissipation occurs only in the vicinity of the quasicrossing points. Fig.~\ref{Fig:Reset}(c) is a part of Fig.~\ref{Fig:Reset}(a) that shows the dynamics of occupations of intermediate levels (Nos.~5, 4, 3, 2) due to dissipation after the quasicrossing point between levels 6 and 7. This subfigure is obtained by continuing to the left and stretching the area bounded by the red dashed rectangle in Fig.~\ref{Fig:Reset}(b) along the ordinate axis.

The dependence of the occupations $\rho_{kk}$ of the qudit levels on $x_\text{e}$ can be approximately calculated: (i)~by the adiabatic-impulse model in the absence of dissipation (see Appendix~\ref{Sec:AppendixA1}) and (ii)~by combination of the adiabatic-impulse model and the rate equation approach in the presence of dissipation (see Appendix~\ref{Sec:AppendixA2}). The approximate method presented in Appendix~\ref{Sec:AppendixA2} gives $x_\text{e}$-dependence of $\rho_{kk}$ which, for chosen parameters, is in a sufficiently good agreement (excluding the vicinities of the quasicrossing points) with the results obtained by numerically solving the system of equations~(\ref{system-reset}) and shown in Fig.~\ref{Fig:Reset}, despite the fact that this approximation does not take into account the specificity of the problem, e.g., the form of the potential $U(x)$ in Eq.~(\ref{potential}).

\section{Conclusions}
\label{Sec:Conclusions}

In summary, we have considered photon detector based on the flux qubit as a multilevel system. Treating electromagnetic field semiclassically and numerically solving the Lindblad equation, we have calculated the dynamics of the occupations of the levels for the readout stage of the detection and obtained that the form of signal does not enter the results for approximations used. We can conclude from the plot showing the dynamics for the readout stage that an input signal leads to a change of the magnetic flux direction. Also, we have investigated the reset stage of the detection and calculated the occupations of the levels as functions of the external magnetic flux.

\section*{Acknowledgements}
We would like to thank O.~M.~Bahrova, P.~Febvre, O.~Yu.~Kitsenko, V.~Yu.~Lyakhno, E.~V.~Stolyarov, and O.~G.~Turutanov for fruitful discussions and help with this work and S.~Ashhab for critically reading this manuscript. 
We acknowledge the support from the IEEE program ``Magnetism for Ukraine 2023". 
O.A.I. was partially supported by NATO Science for Peace and Security Program (Project G5796) and by Grant of the NAS of Ukraine for young scientists ``Mesoscopic systems for quantum interferometry and detection of single photons" (Grant Number 0123U103073).

\appendix

\section{Adiabatic-impulse model and rate equation approach for a multilevel quantum system}
\label{Sec:AppendixA}

\subsection{Simplified adiabatic-impulse model without relaxation}
\label{Sec:AppendixA1}

In the absence of relaxation, the dynamics of the reset stage of the detection, illustrated in Fig.~\ref{Fig:Reset}(a), 
can also be obtained by an approximate analytical method, the adiabatic-impulse model (see Refs.~\cite{ivakhnenko, ryzhov}).

In this model, the whole dynamics is considered as a series of periods of the adiabatic evolution
when the occupations of the energy levels are constant and the diabatic transitions, occurring
at the passages of the adiabatic energy-level quasicrossings with the change of the level occupations.
Generally, in the adiabatic-impulse model, one calculates the matrices of these diabatic transitions and matrices of the adiabatic evolution, and the matrix of the evolution for the whole dynamics is calculated as a multiplication of them.

We consider a particular case, namely, (i)~energy levels are as in Fig.~\ref{Fig:E_Phi-e}(a), (ii)~the external magnetic flux $\Phi_\text{e}$ linearly depends on time [see Eq.~(\ref{Phi-e_reset})], (iii)~only the  lowest energy level is populated in the initial state. In this case, there are passages of quasicrossings one by one [see the dashed arrow in Fig.~\ref{Fig:E_Phi-e}(a)] with no interference and the occupations of the adiabatic energy levels during the reset can also be found using the following simple consideration.
The occupations of the energy levels are constant during the dynamics far from the quasicrossings. During the passage of the $n$-th quasicrossing on the path shown by the dashed arrow in Fig.~\ref{Fig:E_Phi-e}(a) with the minimal splitting $\Delta_n$ between
the adiabatic energy levels $E_n$ and $E_{n+1}$, the probability of  tunneling to the upper energy level $E_{n+1}$ equals 
\begin{equation}\label{final_occupation_n}
	\mathcal{P}_n = \exp[-\pi \Delta^2_n / 2 \hbar v]
\end{equation}
[see Eq.~(\ref{final_occupation})] and the probability of remaining on the same energy level $E_n$ equals $1-\mathcal{P}_n$. This allows to find the occupations of all energy levels for all adiabatic evolution intervals during the reset process. In particular, for the final state the occupation probabilities are given by (see also Ref.~\cite{ashhab2023}):
\begin{equation}
	\begin{aligned}
		& \rho_{11} = 1-\mathcal{P}_1, \\
		& \rho_{22} = \mathcal{P}_1 (1-\mathcal{P}_2), \\
		& \rho_{kk} = \prod_{n=1}^{k-1}\mathcal{P}_n (1-\mathcal{P}_k), \qquad k = 3, ..., 6, \\
		& \rho_{77} = \prod_{n=1}^{6}\mathcal{P}_n.
	\end{aligned}
\end{equation}

\begin{figure}
	\includegraphics[width=7.8cm]{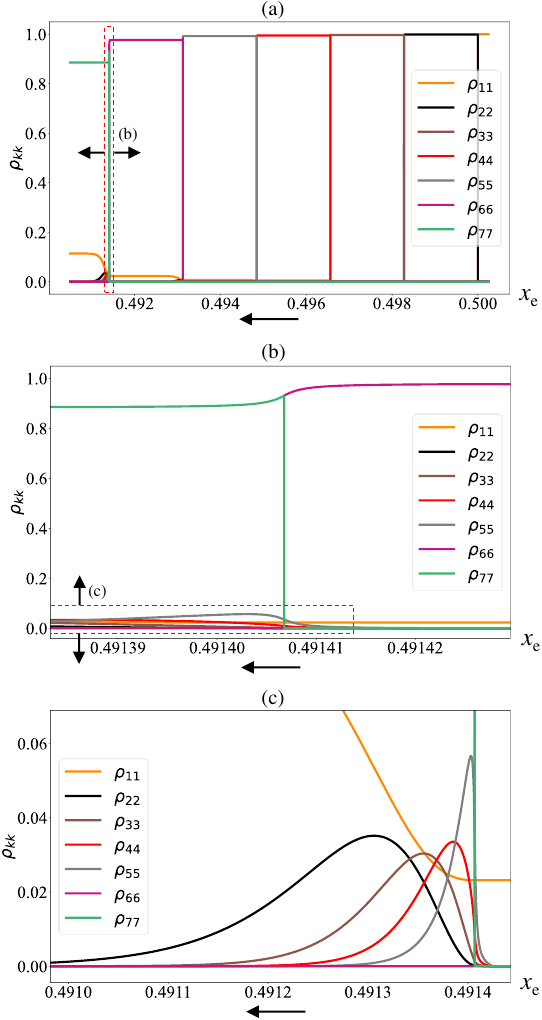}
	\caption{(a)~The dependence of the occupation probabilities $\rho_{kk}$ of the qudit levels on the dimensionless external magnetic flux $x_\text{e}$ during the reset stage of the detection obtained by using the adiabatic-impulse model and the rate equation approach. The  energy splittings obtained by numerically solving the Schr\"odinger equation~(\ref{Schroedinger}) are as follows:
			$\Delta_1=3 \times 10^{-8}$~K, 
			$\Delta_2=7 \times 10^{-5}$~K, 
			$\Delta_3=4 \times 10^{-4}$~K, 
			$\Delta_4=0.001$~K, 
			$\Delta_5=0.002$~K, 
			$\Delta_6=0.003$~K.
			All parameters are the same as in Fig.~\ref{Fig:Reset}(a).
			(b)~The same dependence as in~(a) near the point of quasicrossing of levels 6 (magenta) and 7 (green). (c)~The same dependence as in~(a) that shows the dynamics of occupations of intermediate levels (Nos.~5, 4, 3, 2) due to dissipation after the last quasicrossing point (zoom-in near the $x$-axis).}
	\label{Fig:Reset_rateeq}
\end{figure}

\subsection{Rate equation approach}
\label{Sec:AppendixA2}

In the presence of relaxation, the considered simplified adiabatic-impulse model can be combined with the rate equation approach (see Ref.~\cite{liul}).

On each interval of the adiabatic evolution, we solve numerically the system of rate equations
\begin{equation}\label{system-rate}
	\begin{aligned}
		& \frac{d \rho_{11}}{d t} = \sum_{j=2}^7 \Gamma_{j1} (t) \rho_{jj},	\\
		& \frac{d \rho_{kk}}{d t} = - \sum_{j=1}^{k-1} \Gamma_{kj} (t) \rho_{kk} + \sum_{j=k+1}^7 \Gamma_{jk} (t) \rho_{jj}, 	\ k=2,...,6, \\
		& \frac{d \rho_{77}}{d t} = - \sum_{j=1}^6 \Gamma_{7j} (t) \rho_{77},
	\end{aligned}
\end{equation}
where the time-dependent energy relaxation rate $\Gamma_{jk} (t)$ from the energy level $E_j$ to the energy level $E_k$ ($j>k$) is given by Eq.~(\ref{relaxation-rates}) [see also the sentence below Eq.~(\ref{B})]. In order to obtain the levels occupation probabilities $\rho_{kk}$ as functions of the dimensionless external magnetic flux $x_\text{e}$, we note that the linear dependence between $t$ and $x_\text{e}$ is defined by Eq.~(\ref{Phi-e_reset}). In what follows, we denote by $t_{n,n+1}$ the moment of time corresponding to the quasicrossing point between level $E_n$ and level $E_{n+1}$ (the transition from level $E_n$ to level $E_{n+1}$ is assumed to occur instantaneously in this approximate method). We also denote by $I_1$ the interval between the initial moment of time $t_0=0$ [which corresponds to $\Phi_\text{e}=\Phi_\text{e0}$; see Eq.~(\ref{Phi-e_reset}) and caption to Fig.~\ref{Fig:Reset}(a)] and the moment $t_{12}$, by $I_n$ the interval between $t_{n-1,n}$ and $t_{n,n+1}$ $(n=2,3,...,6)$, and by $I_7$ the interval between $t_{67}$ and the moment $\tau>t_{67}$. Accordingly, $\rho_{kk}^{(n)}(t)$ denotes the occupation probability of level $k$ for $t\in I_n$ $(n=1,2,...,7)$. The initial condition for the probabilities in the interval $I_1$ is $\rho_{11}^{(1)}(0)=1$ and $\rho_{kk}^{(1)}(0)=0$ for $k=2,3,...,7$.

There is no change in the occupation probabilities till the point $t_{12}$, therefore, $\rho_{11}^{(1)}(t_{12})=1$ and $\rho_{kk}^{(1)}(t_{12})=0$ $(k=2,3,...,7)$. Taking into account that at the point $t_{12}$, there is a transition from level $E_1$ to level $E_2$ with the probability $\mathcal{P}_1$ [see Eq.~(\ref{final_occupation_n})], we obtain that the initial condition for the probabilities in the interval $I_2$ is $\rho_{11}^{(2)}(t_{12})=1-\mathcal{P}_1$, $\rho_{22}^{(2)}(t_{12})=\mathcal{P}_1$, and $\rho_{kk}^{(2)}(t_{12})=0$ $(k=3,4,...,7)$.
 
The solution of the system~(\ref{system-rate}) of the rate equations in the interval $I_2$ is obtained numerically. At the point $t_{23}$, there is an LZSM transition from level 2 to level 3 with the probability $\mathcal{P}_2$. Therefore, the initial condition for the probabilities in the interval $I_3$ is $\rho_{11}^{(3)}(t_{23})=\rho_{11}^{(2)}(t_{23})$, $\rho_{22}^{(3)}(t_{23})=(1-\mathcal{P}_2) \rho_{22}^{(2)}(t_{23})$, $\rho_{33}^{(3)}(t_{23})=\mathcal{P}_2 \rho_{22}^{(2)}(t_{23})$, and $\rho_{kk}^{(3)}(t_{23})=0$ $(k=4,5,6,7)$. The system~(\ref{system-rate}) of the rate equations is solved on all subsequent intervals analogously.

For the reset, the dependence of the levels occupations on the dimensionless external magnetic flux obtained by the approximate rate equation approach is illustrated in Fig.~\ref{Fig:Reset_rateeq}, and it is in a good agreement (excluding the vicinities of the quasicrossing points) with the dependence in Fig.~\ref{Fig:Reset} obtained by solving the system of equations~\eqref{system-reset} for the matrix elements of the density operator. We note that according to the solution of the Lindblad equation~\eqref{system-reset}, the LZSM transitions occur during a finite (yet rather small) time interval [see Fig.~\ref{Fig:Reset}(b)], while in the approximate approach based on the adiabatic-impulse model and rate equations, the LZSM transitions are instantaneous [see Fig.~\ref{Fig:Reset_rateeq}(b)]. That leads to discrepancy of the results shown in Figs.~\ref{Fig:Reset} and \ref{Fig:Reset_rateeq}. 

The approximate approach (rate equation method combined with the adiabatic-impulse model) can be used for the description of the problem with arbitrary values $\mathcal{P}_n$ of the LZSM probabilities ($n=1,...,6$). Note, however, that for the parameters $v_\text{e}$ (the rate of change of the external magnetic flux) and $\Delta_n$ (energy splittings) considered in the main text, the LZSM probabilities $\mathcal{P}_n$, calculated by Eq.~\eqref{final_occupation_n}, are equal to unity with high accuracy. Therefore, the fact that $\rho_{n+1,n+1}^{(n+1)}(t'')<\rho_{nn}^{(n)}(t')$, where $t'\lesssim t_{n,n+1}\lesssim t''$, is explained by the presence of the relaxation processes in the vicinity of the quasicrossing point, and not by non-unity LZSM transition probability.


\begin{thebibliography}{99}
	
\bibitem{opremcak2018}
A.~Opremcak, I.~V.~Pechenezhskiy, C.~Howington, B.~G.~Christensen, M.~A.~Beck, E.~Leonard Jr., J.~Suttle, C.~Wilen, K.~N.~Nesterov, G.~J.~Ribeill, T.~Thorbeck, F.~Schlenker, M.~G.~Vavilov, B.~L.~T.~Plourde, R.~McDermott, Measurement of a superconducting qubit with a microwave photon counter, Science {\bf 361}, 1239 (2018).
	
\bibitem{opremcak2021}
A.~Opremcak, C.~H.~Liu, C.~Wilen, K.~Okubo, B.~G.~Christensen, D.~Sank, T.~C.~White, A.~Vainsencher, M.~Guistina, A.~Megrant, B.~Burkett, B.~L.~T.~Plourde, and R.~McDermott, High-fidelity measurement of a superconducting qubit using an on-chip microwave photon counter, Phys. Rev. X {\bf 11}, 011027 (2021).

\bibitem{campagne-ibarcq}
P.~Campagne-Ibarcq, E.~Zalys-Geller, A.~Narla, S.~Shankar, P.~Reinhold, L.~Burkhart, C.~Axline, W.~Pfaff, L.~Frunzio, R.~J.~Schoelkopf, and M.~H.~Devoret, Deterministic remote entanglement of superconducting circuits through microwave two-photon transitions, Phys. Rev. Lett. {\bf 120}, 200501 (2018).

\bibitem{morris}
P.~A.~Morris, R.~S.~Aspden, J.~E.~C.~Bell, R.~W.~Boyd, and M.~J.~Padgett, 
Imaging with a small number of photons, Nat. Commun. {\bf 6}, 5913 (2015).

\bibitem{irastorza} I.~G.~Irastorza and J.~Redondo, New experimental approaches in the search for axion-like particles, Prog. Part. Nucl. Phys. {\bf 102}, 89 (2018).  

\bibitem{ghirri} A.~Ghirri, S.~Cornia, and M.~Affronte, Microwave photon detectors based on semiconducting double quantum dots, Sensors {\bf 20}, 4010 (2020).

\bibitem{gu} 
X.~Gu, A.~F.~Kockum, A.~Miranowicz, Y.-x.~Liu, and F.~Nori, Microwave photonics with superconducting quantum circuits, Phys. Rep. {\bf 718-719}, 1 (2017).

\bibitem{kono}
S.~Kono, K.~Koshino, Y.~Tabuchi, A.~Noguchi, and Y.~Nakamura, Quantum non-demolition detection of an itinerant microwave photon, Nat. Phys. {\bf 14}, 546 (2018).

\bibitem{besse2018}
J.-C.~Besse, S.~Gasparinetti, M.~C.~Collodo, T.~Walter, P.~Kurpiers, M.~Pechal, C.~Eichler, and A.~Wallraff, Single-shot quantum nondemolition detection of individual itinerant microwave photons, Phys. Rev. X {\bf 8}, 021003 (2018).

\bibitem{chen}
Y.-F.~Chen, D.~Hover, S.~Sendelbach, L.~Maurer, S.~T.~Merkel, E.~J.~Pritchett, F.~K.~Wilhelm, and R.~McDermott, Microwave photon counter based on Josephson junctions, 
Phys. Rev. Lett. {\bf 107}, 217401 (2011).

\bibitem{inomata}
K.~Inomata, Z.~Lin, K.~Koshino, W.~D.~Oliver, J.-S.~Tsai, T.~Yamamoto, and Y.~Nakamura, Single microwave-photon detection using an artificial $\Lambda$-type three-level system, Nat. Commun. {\bf 7}, 12303 (2016).

\bibitem{govia}
L.~C.~G.~Govia, E.~J.~Pritchett, S.~T.~Merkel, D.~Pineau, F.~K.~Wilhelm, Theory of Josephson photomultipliers: Optimal working conditions and back action, Phys. Rev. A {\bf 86}, 032311 (2012).

\bibitem{semenov}
E.~V.~Stolyarov, O.~V.~Kliushnichenko, V.~S.~Kovtoniuk, and A.~A.~Semenov, Photon-number resolution with microwave Josephson photomultipliers, Phys. Rev. A {\bf 108}, 063710 (2023).

\bibitem{koshino}
K.~Koshino, K.~Inomata, Z.~Lin, Y.~Nakamura, and T.~Yamamoto, 
Theory of microwave single-photon detection using an impedance-matched $\Lambda$ system, Phys. Rev. A {\bf 91}, 043805 (2015).

\bibitem{shapovalov}
A.~P.~Shapovalov, V.~E.~Shaternik, O.~G.~Turutanov, V.~Yu.~Lyakhno, V.~I.~Shnyrkov, 
On the possibility of faster detection of magnetic flux changes in a single-photon counter by RF SQUID with MoRe–Si(W)–MoRe junction, 
Low Temp. Phys. {\bf 45}, 776 (2019).

\bibitem{lyakhno}
V.~Yu.~Lyakhno, O.~G.~Turutanov, A.~P.~Boichenko, A.~P.~Shapovalov, A.~A.~Kalenyuk, V.~I.~Shnyrkov, 
Hybrid shield for microwave single-photon counter based on a flux qubit, 
Low Temp. Phys. {\bf 48}, 228 (2022).

\bibitem{shnyrkov2020}
V.~I.~Shnyrkov, Wu~Yangcao, O.~G.~Turutanov, V.~Yu.~Lyakhno, and A.~A.~Soroka, Scheme for flux-qubit-based microwave single-photon counter with weak continuous measurement, 2020 IEEE Ukrainian Microwave Week, p.~737.

\bibitem{schmidt}
V.~V.~Schmidt, {\it The physics of superconductors. Introduction to fundamentals and applications}, Springer (1997), p.~93.

\bibitem{shnyrkov}
V.~I.~Shnyrkov, Wu~Yangcao, A.~A.~Soroka, O.~G.~Turutanov, and V.~Yu.~Lyakhno, Frequency-tuned microwave-photon counter based on a superconductive quantum interferometer, 
Low Temp. Phys. {\bf 44}, 213 (2018).

\bibitem{ashhab} 
S.~Ashhab, O.~A.~Ilinskaya, and S.~N.~Shevchenko, Nonlinear Landau-Zener-St\"uckelberg-Majorana problem, Phys. Rev. A {\bf 106}, 062613 (2022).

\bibitem{kofman}
P.~O.~Kofman, S.~N.~Shevchenko, and F.~Nori, Tuning the initial phase to control the final state of a driven qubit, Phys. Rev. A {\bf 109}, 022409 (2024).

\bibitem{ivakhnenko}
O.~V.~Ivakhnenko, S.~N.~Shevchenko, and F.~Nori, Nonadiabatic Landau-Zener-St\"uckelberg-Majorana transitions, dynamics, and interference, Phys. Rep. {\bf 995}, 1 (2023).

\bibitem{omelyanchouk}
A.~N.~Omelyanchouk, S.~N.~Shevchenko, Ya.~S.~Greenberg, O.~Astafiev, and E.~Il'ichev, Quantum behavior of a flux qubit coupled to a resonator, Low Temp. Phys. {\bf 36},  
893 (2010).

\bibitem{govia2014}
L.~C.~G.~Govia, E.~J.~Pritchett, C.~Xu, B.~L.~T.~Plourde, M.~G.~Vavilov, F.~K.~Wilhelm, and R.~McDermott, High-fidelity qubit measurement with a microwave-photon counter, Phys. Rev. A {\bf 90}, 062307 (2014).

\bibitem{jones}
P.~J.~Jones, J.~Salmilehto, M.~M\"ott\"onen, Highly controllable qubit-bath coupling based on a sequence of resonators, J. Low Temp. Phys. {\bf 173}, 152 (2013).	

\bibitem{makhlin}
Y.~Makhlin, G.~Sch\"on, A.~Shnirman, Quantum-state engineering with Josephson-junction devices, Rev. Mod. Phys. {\bf 73}, 357 (2001).

\bibitem{kumar}
P.~Kumar, S.~Sendelbach, M.~A.~Beck, J.~W.~Freeland, Z.~Wang, H.~Wang, C.~C.~Yu, R.~Q.~Wu, D.~P.~Pappas, and R.~McDermott, Origin and reduction of $1/f$ magnetic flux noise in superconducting devices, Phys. Rev. Appl. {\bf 6}, 041001 (2016).

\bibitem{yoshihara2006}
F.~Yoshihara, K.~Harrabi, A.~O.~Niskanen, Y.~Nakamura, and J.~S.~Tsai, Decoherence of flux qubits due to $1/f$ flux noise, Phys. Rev. Lett. {\bf 97}, 167001 (2006).

\bibitem{rower}
D.~A.~Rower, L.~Ateshian, L.~H.~Li, M.~Hays, D.~Bluvstein, L.~Ding, B.~Kannan, A.~Almanakly, J.~Braum\"uller, D.~K.~Kim, A.~Melville, B.~M.~Niedzielski, M.~E.~Schwartz, J.~L.~Yoder, T.~P.~Orlando, J.~I-J.~Wang, S.~Gustavsson, J.~A.~Grover, K.~Serniak, R.~Comin, and W.~D.~Oliver, Evolution of $1/f$ flux noise in superconducting qubits with weak magnetic fields, Phys. Rev. Lett. {\bf 130}, 220602 (2023).

\bibitem{brookes}
P.~Brookes, G.~Tancredi, A.~D.~Patterson, J.~Rahamim, M.~Esposito, T.~K.~Mavrogordatos, P.~J.~Leek, E.~Ginossar, M.~H.~Szymanska, 
Critical slowing down in circuit quantum electrodynamics, 
Sci. Adv. {\bf 7}, eabe9492 (2021).

\bibitem{rabi}
I.~I.~Rabi, Space quantization in a gyrating magnetic field, Phys. Rev. {\bf 51}, 652 (1937).

\bibitem{jaynes-cummings}
E.~T.~Jaynes and F.~W.~Cummings, Comparison of quantum and semiclassical radiation theories with application to the beam maser, Proc. IEEE {\bf 51}, 89 (1963).

\bibitem{yoshihara2022}
F.~Yoshihara, S.~Ashhab, T.~Fuse, M.~Bamba, and K.~Semba, Hamiltonian of a flux-qubit-LC oscillator circuit in the deep--strong-coupling regime, Sci. Rep. {\bf 12}, 6764 (2022).

\bibitem{yoshihara2017}
F.~Yoshihara, T.~Fuse, S.~Ashhab, K.~Kakuyanagi, S.~Saito, and K.~Semba, 
Superconducting qubit--oscillator circuit beyond the ultrastrong-coupling regime, Nat. Phys. {\bf 13}, 44 (2017).

\bibitem{baker}
B.~Baker, A.~C.~Y.~Li, N.~Irons, N.~Earnest, and J.~Koch, Adaptive rotating-wave approximation for driven open quantum systems, Phys. Rev. A {\bf 98}, 052111 (2018).

\bibitem{shevchenko}
S.~N.~Shevchenko, Mesoscopic physics meets quantum engineering, World Scientific (2019).

\bibitem{ryzhov}
A. I. Ryzhov, O. V. Ivakhnenko, S. N. Shevchenko, M. F. Gonzalez-Zalba, Franco Nori, Alternative fast quantum logic gates using nonadiabatic Landau-Zener-St\"{u}ckelberg-Majorana transitions, arXiv:2310.17932.

\bibitem{ashhab2023}
S.~Ashhab, T.~Fuse, F.~Yoshihara, S.~Kim, and K.~Semba, Controlling qubit-oscillator systems using linear parameter sweeps, New J. Phys. {\bf 25}, 093011 (2023).

\bibitem{liul}
M. P. Liul, A. I. Ryzhov, and S. N. Shevchenko, Interferometry of multi-level systems: rate-equation approach for a charge qudit, Eur. Phys. J. Spec. Top. {\bf 232}, 3227 (2023).



\end{thebibliography}
\end{document}